%% file: main.tex
\documentclass[%
  copyright,creativecommons,
]{eptcs}
\providecommand{\event}{PLACES 2017} 
\usepackage{underscore}           

\usepackage[british]{babel}

\input{packages.tex}%
\input{macros.tex}%

\title{%
  Multiparty Session Types, Beyond Duality\\
  {(Abstract)}
}%

\author{%
  Alceste Scalas%
  \institute{Imperial College London}%
  \email{alceste.scalas@imperial.ac.uk}%
  \and%
  Nobuko Yoshida%
  \institute{Imperial College London}%
  \email{n.yoshida@imperial.ac.uk}%
}
\def\titlerunning{Multiparty Session Types, Beyond Duality (Abstract)}
\def\authorrunning{A.~Scalas and N.~Yoshida}
\begin{document}
\maketitle

\input{contents.tex}%

{\small%
  \bibliographystyle{eptcsini}%
  \bibliography{main}%
}%


\end{document}

%% file: packages.tex
\usepackage{etoolbox}

\usepackage{amsmath}
\usepackage{amsthm}

\usepackage{amssymb}

\usepackage{cmll}
\usepackage{colonequals}

\usepackage[inline]{enumitem}%

\usepackage{fixltx2e}

\usepackage{import}

\usepackage{lipsum}

\usepackage{MnSymbol}

\usepackage{multibib}
\newcites{app}{Additional references for the Appendix}

\usepackage{nicefrac}

\usepackage{pifont}

\usepackage{proof}

\usepackage{stmaryrd}

\usepackage{thmtools}
\usepackage{thm-restate}

\usepackage[usenames,dvipsnames]{xcolor}

\usepackage{tikz}
\usetikzlibrary{arrows,shadows}
\usepackage{pgf-umlsd}

\usepackage{url}

\usepackage{xifthen}
\usepackage{xspace}

\usepackage{wrapfig}

\usepackage{hyperref} 
\hypersetup{final} 

\usepackage{mathtools} 

\usepackage{cleveref}

%% file: macros.tex
\newif\ifdraft%
  \draftfalse%
  \drafttrue%

\newcommand{\ifempty}[3]{%
  \ifthenelse{\isempty{#1}}{#2}{#3}%
}%

\def\cf{cf.\@\xspace}%
\def\eg{e.g.\@\xspace}%
\def\Eg{E.g.\@\xspace}%
\def\wrt{w.r.t.\@\xspace}%

\definecolor{ruleColor}{rgb}{0.1, 0.3, 0.1}
\newcommand{\inferrule}[1]{{\color{ruleColor}\textsc{\scriptsize [#1]}}}%
\definecolor{groundColor}{rgb}{0.38, 0.25, 0.32}

\definecolor{roleColor}{rgb}{0.5, 0.0, 0.0}
\newcommand{\roleFmt}[1]{{\color{roleColor}\mathtt{#1}}}%
\newcommand{\roleP}[1][]{%
  \ifempty{#1}{\roleFmt{p}}{\roleFmt{p}_{{\color{roleColor}#1}}}%
}%
\newcommand{\roleQ}[1][]{%
  \ifempty{#1}{\roleFmt{q}}{\roleFmt{q}_{{\color{roleColor}#1}}}%
}%
\newcommand{\roleR}[1][]{%
  \ifempty{#1}{\roleFmt{r}}{\roleFmt{r_{{\color{roleColor}#1}}}}%
}%

\newcommand{\labFmt}[2][]{\ifempty{#1}{\mathtt{#2}}{\mathtt{#2}\textsubscript{#1}}}%

\definecolor{mpColor}{rgb}{0, 0, 0}
\newcommand{\mpFmt}[1]{{\color{mpColor}#1}}%

\newcommand{\mpChanRole}[2]{\mpFmt{{#1}[{#2}]}}%

\newcommand{\mpPar}{\mathbin{\mpFmt{\mid}}}%
\newcommand{\mpBigPar}[2]{\mathbin{\mpFmt{\big\mid_{#1}}{#2}}}%
\newcommand{\mpRes}[2]{\mpFmt{\left(\mathbf{\nu}{#1}\right){#2}}}%
\newcommand{\mpS}[1][]{\mpFmt{\ifempty{#1}{s}{s_{#1}}}}%
\newcommand{\mpP}[1][]{\mpFmt{\ifempty{#1}{P}{P_{#1}}}}%
\newcommand{\mpPi}[1][]{\mpFmt{\ifempty{#1}{P'}{P'_{#1}}}}%
\newcommand{\mpMove}{\to}%
\newcommand{\mpMoveStar}{\mathrel{\mpMove{}^{\!\!\!\!*}}}%
\definecolor{gtColor}{rgb}{0.43, 0.21, 0.1}
\newcommand{\gtFmt}[1]{{\color{gtColor}#1}}%
\newcommand{\gtMsgFmt}[1]{\gtFmt{\labFmt{#1}}}%

\newcommand{\gtG}[1][]{\gtFmt{\ifempty{#1}{G}{G_{#1}}}}%

\newcommand{\gtSeq}{\mathbin{\gtFmt{.}}}%

\newcommand{\gtCommRaw}[3]{%
  \gtFmt{%
    {#1} {\to} {#2}{:}%
    \left\{%
      {#3}%
    \right\}%
  }%
}%
\newcommand{\gtCommSingle}[5]{%
  \gtFmt{%
    {#1} {\to} {#2}{:}%
    \gtCommChoice{#3}{#4}{#5}%
  }%
}%
\newcommand{\gtCommChoice}[3]{%
  \gtFmt{%
    \gtMsgFmt{#1}\ifempty{#2}{}{({#2})}%
    \ifempty{#3}{}{\vphantom{x} \gtSeq {#3}}%
  }%
}%

\newcommand{\gtEnd}{\gtFmt{\mathbf{end}}}%

\definecolor{stColor}{rgb}{0, 0, 0.9}
\newcommand{\stFmt}[1]{{\color{stColor}#1}}%

\newcommand{\stTypeBool}{\stFmt{\operatorname{Bool}}}%
\newcommand{\stTypeInt}{\stFmt{\operatorname{Int}}}%
\newcommand{\stTypeString}{\stFmt{\operatorname{Str}}}%

\newcommand{\stIn}[3]{\ifempty{#1}{}{\roleFmt{#1}}\stFmt{\&#2\ifempty{#3}{}{({#3})}}}%
\newcommand{\stOut}[3]{\ifempty{#1}{}{\roleFmt{#1}}\stFmt{\oplus#2\ifempty{#3}{}{({#3})}}}%

\newcommand{\stChoice}[2]{\stFmt{#1}\ifempty{#2}{}{\stFmt{({#2})}}}%

\newcommand{\stSeq}{\mathbin{\!\stFmt{.}\!}}%
\newcommand{\stIntC}{\mathbin{\stFmt{\oplus}}}%
\newcommand{\stIntSum}[3]{\roleFmt{#1}\,\stFmt{\bigoplus_{#2\!}{#3}}}%
\newcommand{\stExtC}{\mathbin{\stFmt{\&}}}%
\newcommand{\stExtSum}[3]{\roleFmt{#1}\,\stFmt{\bigwith_{#2\!}{#3}}}%
\newcommand{\stLabFmt}[1]{\stFmt{\labFmt{#1}}}%

\newcommand{\stS}[1][]{\stFmt{\ifempty{#1}{S}{S_{#1}}}}%
\definecolor{ptColor}{rgb}{0.20, 0.29, 0.09}
\newcommand{\stEnv}[1][]{\stFmt{\ifempty{#1}{\Gamma}{\Gamma_{\!#1}}}}%
\newcommand{\stEnvi}[1][]{\stFmt{\ifempty{#1}{\Gamma'}{\Gamma'_{\!#1}}}}%
\newcommand{\stEnvMap}[2]{\stFmt{\mpFmt{#1}\mathbin{\!:\!}{#2}}}%
\newcommand{\stEnvComp}{\mathpunct{\stFmt{,}}}%
\newcommand{\stEnvGR}[2]{\stFmt{#1} \triangleleft \stFmt{#2}}

\newcommand{\mpEnv}[1][]{\stFmt{\ifempty{#1}{\Theta}{\Theta_{#1}}}}%
\newcommand{\stJudge}[4]{%
  \stFmt{{#1} \vdash \mpFmt{#4} \triangleright \stEnvGR{#2}{#3}}%
}%

\newcommand{\stJudgeStd}[3]{%
  \stFmt{{#1} \vdash \mpFmt{#3} \triangleright {#2}}%
}%

\newcommand{\negSpace}[1][]{\ifempty{#1}{\vspace{-1mm}}{\vspace{-#1mm}}}%

%% file: contents.tex
Multiparty Session Types (MPST) %
are a well-established typing discipline for message-passing processes %
interacting on \emph{sessions} involving two or more participants. %
Session typing can ensure desirable properties:
absence of communication errors and deadlocks, and protocol conformance. %
However, existing MPST works %
provide a \emph{subject reduction} result %
that is arguably (and sometimes, surprisingly) restrictive: %
it only holds for 
typing contexts with strong \emph{duality} constraints %
on the interactions between pairs of participants. %
Consequently, many ``intuitively correct'' examples %
cannot be typed and/or cannot be proved type-safe.
We illustrate some of these examples, %
and discuss the reason for these limitations.
Then, we outline a novel MPST typing system %
that removes these restrictions.

\paragraph{MPST in a Nutshell}%

\newcommand{\gtExChoice}{\gtG}%

In the MPST framework \cite{HYC08}, %
\emph{global types} (describing interactions among %
\emph{roles}) %
are projected to \emph{local types} %
used to type-check \emph{processes}. %
\Eg, the %
global type $\gtExChoice$ %
involves roles %
$\roleP$, $\roleQ$, $\roleR$: %
\begin{equation*}
  \label{eq:gtexchoice}%
    \gtExChoice \;=\;%
    \gtCommRaw{\roleP}{\roleQ}{\!\!\!\!%
      \begin{array}{l}%
        \gtCommChoice{m1}{\stTypeInt}{%
          \gtCommSingle{\roleQ}{\roleR}{m2}{\stTypeString}{%
            \gtCommSingle{\roleR}{\roleP}{m3}{\stTypeBool}{%
              \gtEnd%
            }%
          }%
        }\,,%
        \\%
        \gtCommChoice{stop}{}{%
          \gtCommSingle{\roleQ}{\roleR}{quit}{}{%
              \gtEnd%
          }%
        }%
      \end{array}
      \!\!\!\!%
    }%
\end{equation*}

\noindent%
$\gtG$ says that %
$\roleP$ sends to $\roleQ$ \emph{either} a message $\gtMsgFmt{m1}$ %
(carrying an $\stTypeInt$) %
\emph{or} $\gtMsgFmt{stop}$; %
in the first case, %
$\roleQ$ sends $\gtMsgFmt{m2}$ to $\roleR$ %
(carrying a $\stTypeString$), %
then $\roleR$ sends $\gtMsgFmt{m3}$ to $\roleP$ %
(carrying a $\stTypeBool$), %
and the session $\gtEnd$s; %
otherwise, %
in the second case, $\roleQ$ sends $\gtMsgFmt{quit}$ to $\roleR$, %
and the session $\gtEnd$s.
The \emph{projections of $\gtG$} %
are the I/O actions of each role in $\gtExChoice$:
\newcommand{\stExChoiceP}{\stS[\roleP]}%
\newcommand{\stExChoiceQ}{\stS[\roleQ]}%
\newcommand{\stExChoiceR}{\stS[\roleR]}%
\begin{equation*}
  \label{eq:gtexchoice-proj}%
  \small%
  \hspace{-0.5mm}%
      \stExChoiceP =%
      \stIntSum{\roleQ}{}{\left\{%
        \begin{array}{@{\hskip 0mm}l@{\hskip 0mm}}%
          \stChoice{\stLabFmt{m1}}{\stTypeInt} \,\stSeq\,%
          \stIn{\roleR}{%
            \stChoice{\stLabFmt{m3}}{\stTypeBool}
          }{}%
          \,,%
          \\%
          \stChoice{\stLabFmt{stop}}{}
        \end{array}
      \right\}}%
      \quad%
      \stExChoiceQ =%
      \stExtSum{\roleP}{}{\left\{%
        \begin{array}{@{\hskip 0mm}l@{\hskip 0mm}}%
          \stChoice{\stLabFmt{m1}}{\stTypeInt} \,\stSeq\,%
          \stOut{\roleR}{%
            \stChoice{\stLabFmt{m2}}{\stTypeString}
          }{}%
          \,,%
          \\%
          \stChoice{\stLabFmt{stop}}{} \,\stSeq\,%
          \stOut{\roleR}{%
            \stChoice{\stLabFmt{quit}}{}
          }{}%
        \end{array}
      \right\}}%
      \quad%
      \stExChoiceR =%
      \stExtSum{\roleQ}{}{\left\{%
        \begin{array}{@{\hskip 0mm}l@{\hskip 0mm}}%
          \stChoice{\stLabFmt{m2}}{\stTypeString} \,\stSeq\,%
          \stOut{\roleP}{%
            \stChoice{\stLabFmt{m3}}{\stTypeBool}
          }{}%
          \,,%
          \\%
          \stChoice{\stLabFmt{quit}}{}
        \end{array}
      \right\}}%
\end{equation*}
Here, %
$\stExChoiceP$, $\stExChoiceQ$, $\stExChoiceR$ %
are the projections of $\gtExChoice$ %
resp.~onto $\roleP$, $\roleQ$, $\roleR$. %
\Eg, %
$\stExChoiceP$ is a session type %
that represents the behaviour of $\roleP$ in $\gtExChoice$: %
it must send ($\stIntC$) to $\roleQ$ either %
$\stChoice{\stLabFmt{m1}}{\stTypeInt}$ %
or $\stChoice{\stLabFmt{stop}}{}$; %
in the first case, %
the channel is then used %
to receive ($\stExtC$) %
message $\stChoice{\stLabFmt{m3}}{\stTypeBool}$ from $\roleR$, %
and the session ends; %
otherwise, %
in the second case, %
the session ends. %
Now, a \emph{typing context $\stEnv$} %
can assign types $\stExChoiceP$, $\stExChoiceQ$ and $\stExChoiceR$ %
to \emph{multiparty channels} %
$\mpChanRole{\mpS}{\roleP}$, $\mpChanRole{\mpS}{\roleQ}$ and %
$\mpChanRole{\mpS}{\roleR}$, %
used to play roles $\roleP$,
$\roleQ$ and $\roleR$ on \emph{session $\mpS$}. %
Then, %
if \eg 
some parallel processes %
$\mpP[\roleP]$, $\mpP[\roleQ]$ and $\mpP[\roleR]$ %
type-check \wrt $\stEnv$, %
then we know that such processes use the channels %
abiding by their types. %

\paragraph{Subject Reduction, or Lack Thereof}%
We would expect %
that typed processes reduce type-safely, %
\eg:
\begin{equation}
  \label{eq:subject-reduction:ideal}%
  \stJudgeStd{}{\stEnv}{\mpP}%
  \;\text{ and }\;%
  \mpP \mpMoveStar \mpPi%
  \quad\text{implies}\quad%
  \exists \stEnvi:\;%
  \stJudgeStd{}{\stEnvi}{\mpPi}%
  \quad%
  \text{\footnotesize%
    (where\; %
    $\mpP = \mpP[\roleP] \mpPar \mpP[\roleQ] \mpPar \mpP[\roleR]$ %
    \;and\; %
    $\stEnv =%
    \stEnvMap{\mpChanRole{\mpS}{\roleP}}{\stExChoiceP}%
    \stEnvComp%
    \stEnvMap{\mpChanRole{\mpS}{\roleQ}}{\stExChoiceQ}%
    \stEnvComp%
    \stEnvMap{\mpChanRole{\mpS}{\roleR}}{\stExChoiceR}$)%
  }%
  \negSpace%
\end{equation}

\noindent%
But surprisingly, \emph{this is not the case!} %
In MPST works (\eg, \cite{Coppo2015GentleIntroMAPST}), %
the subject reduction statement reads:%
\begin{align}
  \label{eq:subject-reduction:std}%
  &\stJudgeStd{}{\stEnv}{\mpP}%
  \;\text{ \emph{\underline{with $\stEnv$ consistent}}}\;%
  \;\;\text{ and }\;\;%
  \mpP \mpMoveStar \mpPi%
  \quad\text{implies}\quad%
  \exists \stEnvi \text{ \emph{\underline{consistent}} such that }\;%
  \stJudgeStd{}{\stEnvi}{\mpPi}%
\end{align}

Intuitively, $\stEnv$ is consistent %
if all its potential interactions between pairs of roles %
are \emph{dual}: %
\eg, all potential outputs of $\stExChoiceP$ towards $\roleR$ %
are matched %
by compatible input capabilities of $\stExChoiceR$ from $\roleP$. %
Consistency %
is quite restrictive, due to its (rather intricate) \emph{syntactic} nature%
---and does \emph{not} hold in our example. %
This is due to %
\emph{inter-role dependencies}: %
$\stExChoiceP$ allows to decide what to send to
$\roleQ$ %
--- and depending on such a choice, %
whether to input $\stLabFmt{m3}$ from $\roleR$, or not. %
This breaks the definition of consistency %
between $\stExChoiceP$ and $\stExChoiceR$; %
hence, $\stEnv$ in \eqref{eq:subject-reduction:ideal} %
is not consistent, %
and %
we cannot apply \eqref{eq:subject-reduction:std} %
to ensure that $\mpP[\roleP]$, $\mpP[\roleQ]$, $\mpP[\roleR]$ %
reduce type-safely. %

\paragraph{Our Proposal}%
In ``standard'' MPST works, %
consistency cannot be lifted %
without breaking subject reduction %
\cite[p.163]{Coppo2015GentleIntroMAPST}. %
Hence, to prove that our example is type-safe, %
we need to revise the MPST foundations. %
We propose a \emph{novel MPST typing system} %
that safely lifts the consistency requirement, 
by introducing:

\begin{enumerate}
\item%
  a new MPST typing judgement with the form\; %
  $\stJudge{\mpEnv}{\stEnv[g]}{\stEnv[r]}{\mpP}$\; %
  ---where $\stEnv[g]$ and $\stEnv[r]$ are respectively %
  the \emph{guarantee} and \emph{rely typing contexts}. %
  Intuitively, %
  $\stEnv[g]$ 
  describes how $\mpP$ uses its channels, %
  while $\stEnv[r]$ %
  describes how other processes %
  (possibly interacting with $\mpP$) %
  are expected to use their channels;
\item%
  a \emph{semantic} notion of typing context safety, %
  called \emph{liveness}, %
  based on MPST context reductions %
  \cite{Coppo2015GentleIntroMAPST}. %
  In our typing judgement, %
  the pair $\stEnv[g] \stEnvComp \stEnv[r]$ must be live: %
  this ensures that 
  each output %
  can synchronise with a compatible input %
  (and \emph{vice versa}). %
  Unlike consistency, %
  liveness supports complex inter-role dependencies, %
  and %
  ensures that the typing context cannot deadlock.
\end{enumerate}

\paragraph{Related Work}%
A technical report %
with more examples and discussion %
is available in \cite{TECHREPORT}. %
Our novel typing system allows to prove type safety of processes %
implementing global types %
with complex inter-role dependencies and delegations. %
To the best of our knowledge, %
the only work %
with a similar capability %
is \cite{DGJPY15Precisenes}; %
however, %
its process calculus only supports \emph{one} session, %
and this restriction is crucially exploited %
to type parallel compositions %
without ``splitting'' them %
(\cf Table~8, rule \inferrule{T-SESS}). %
Hence, unlike our work, \cite{DGJPY15Precisenes} does not support %
multiple sessions and delegation%
---and extending it seems challenging.
Further, %
unlike \cite{DGJPY15Precisenes}, %
our typing rules do \emph{not} depend on global types and projections: %
by removing this orthogonal concern, %
we simplify the theory.
If needed, a set of local types can be related to a global type %
via %
``top-down'' projection %
or ``bottom-up'' synthesis %
\cite{
  Lange2015GraphChor}.
Similarly to most MPST papers, %
our work ensures that a typed process %
$\mpRes{\mpS}{(\mpBigPar{\roleP \in I}{\mpP[\roleP]})}$, %
with each $\mpP[\roleP]$ only interacting on $\mpChanRole{\mpS}{\roleP}$, %
is deadlock-free%
---but does not guarantee deadlock freedom %
for multiple interleaved sessions %
\cite{
CDYP16GlobalProgressMSCS}: %
we leave this topic as future work.%

\smallskip%
{\small%
\noindent%
\textbf{Thanks} %
to the reviewers for their suggestions, %
and to R.\,Hu, J.\,Lange, B.\,Toninho %
for the fruitful discussion. 
Work supported by: %
EPSRC %
(EP/K011715/1,\,EP/K034413/1,\,EP/L00058X/1), %
EU %
(COST\,Action\,IC1201, %
FP7-612985).
}%